\documentclass[aps,prc,twocolumn,superscriptaddress,nofootinbib]{revtex4}
\usepackage[utf8x]{inputenc}
\usepackage{graphicx}
\usepackage[hyperindex=true]{hyperref}
\usepackage{color}
\usepackage{dcolumn}

\newcommand{\Fop}{\hat{F}}

\newcommand{\Qop}{\hat{Q}}

\newcommand{\expec}[1]{\langle #1 \rangle}
\newcommand{\bra}[1]{\langle #1 \vert}
\newcommand{\ket}[1]{\vert #1 \rangle}
\newcommand{\bran}[1]{\langle #1}

\newcommand{\crea}{\hat{a}^\dagger}
\newcommand{\anni}{\hat{a}}

\newcommand{\Ox}[1]{^{#1}\mbox{O}}

\newcommand{\Sn}[1]{^{#1}\mbox{Sn}}

\newcommand{\ljs}   {s_{  1 /2}}
\newcommand{\ljp}[1]{p_{ #1 /2}}

\newcommand{\<}{\langle}
\renewcommand{\>}{\rangle}
\renewcommand{\(}{\left(}
\renewcommand{\)}{\right)}
\newcommand{\hb}{\hbar}

\begin{document}

\title{Structure and direct decay of Giant Monopole Resonances}
\author{B. Avez}\affiliation{Universit\'e Bordeaux 1, CNRS/IN2P3, Centre d'\'Etudes Nucl\'eaires de Bordeaux Gradignan, 
CENBG, Chemin du Solarium, BP120, 33175 Gradignan, France}\affiliation{CEA, Centre de Saclay, IRFU/Service de Physique 
Nucl\'eaire, F-91191 Gif-sur-Yvette, France}
\author{C. Simenel}\affiliation{Department of Nuclear Physics, Research School of Physics and Engineering, Australian National 
University, Canberra, Australian Capital Territory 0200, Australia}\affiliation{CEA, Centre de Saclay, IRFU/Service de Physique Nucl\'eaire, F-91191 Gif-sur-Yvette, France}
\date{\today}

\begin{abstract}
We study structure and direct decay of the Giant Monopole Resonance (GMR) at the RPA level using the Time-Dependent Energy Density Functional 
method in the linear response regime in a few doubly-magic nuclei. 
A proper treatment of the continuum, 
through the use of large coordinate space, allows for a separation between the nucleus  and its emitted nucleons. The 
microscopic structure of the GMR is investigated with the decomposition of the strength function into individual single-particles 
quantum numbers. A similar microscopic decomposition of the spectra of emitted nucleons by direct decay of the GMR is performed.
In this harmonic picture of giant resonance, shifting every contribution by the initial single-particle energy allows to reconstruct the GMR strength function. 
The RPA residual interaction couples bound 1-particle 1-hole states to unbound ones, allowing for the total decay of the GMR. 
In this article, we then intend to get an understanding of the direct decay mechanism from coherent one-particle-one-hole superpositions, while neglecting more complex configurations. 
Time-dependent beyond mean-field approaches should be use, in the future, to extend this method. 
\end{abstract}
\pacs{24.30.Cz, 21.60.Jz}

\maketitle

\section{Introduction}
\label{sec:intro}

Atomic nuclei are known to exhibit a wide range of collective excitations~\cite{Boh75}. 
Amongst them, giant resonances (GRs) are of particular interest in our understanding of collective motion in nuclei. 
Macroscopically, they are associated to high energy modes of small amplitude vibration of the entire nucleus.
Microscopically, they can be modeled, in the harmonic picture, by coherent superpositions of one-particle 
one-hole (1p1h) excitations \cite{Boh75}, although more complex configurations such as 2p2h excitations are necessary to reproduce their experimental widths~\cite{Har01}. 

The width of GRs has three components: the escape width ($\Gamma^\uparrow$), the Landau damping ($\Gamma^L$), and the spreading width ($\Gamma^\downarrow$).
The escape width is due to the GR decay by particle emission. 
Its microscopic origin is the coupling of the correlated $1p1h$ states to the continuum. 
Landau damping occurs due to a one-body coupling to non-coherent $1p1h$ states~\cite{pin66}. 
The spreading width is due to the residual interaction coupling $1p1h$ states to $2p2h$ states.
The $2p2h$ states can also couple to $3p3h$ states and more complex $npnh$ configurations until an equilibrated system is reached. 

The principal approaches to treat collective vibrations in terms of 1p1h constituents are the Random Phase Approximation (RPA)~\cite{boh53} and its relativistic extensions~\cite{lhu89,oak92,vre97}. 
The RPA equations can be obtained as a linearization of the time-dependent Hartree-Fock (TDHF) equation~\cite{Rin80}. 
The recent increase of computational power allowed for the development of realistic TDHF codes able to follow the dynamics of $\sim500$ interacting nucleons \cite{gol09,ked10}.
As a result, several modern TDHF codes have been used to investigate the linear response to vibrational excitations  \cite{Uma05,Mar05,Nak05,fra12} (see also Ref.~\cite{sim12} for a recent review). 
In addition, pairing correlations can be included within the Quasiparticle-RPA (QRPA)~\cite{Rin80,Bla86}, which has been widely used to study various GR strength distributions, either by solving the QRPA equations (see, e.g., Refs.~\cite{eng99,Kha02,kha04,fra05,per08,yos11}), or with small amplitude real time calculations~\cite{Ave08,Eba10,ste11,has12}.

The main drawback of these approaches is the limitation to the independent (quasi)particle picture which prevents the inclusion of more complex configurations. 
In particular, the spreading width cannot be determined at the RPA level because 2p2h configurations are neglected. 
Several approaches have then been considered to go beyond the independent (quasi)particle description of quantum vibrations.
The inclusion of (quasi)particle-phonon coupling \cite{bor81,kam04} has shown to be very efficient to reproduce the width of giant resonances, e.g., in tin isotopes \cite{tse09}. 
The extended-TDHF approach \cite{Lac04}, and the time-dependent density-matrix theory \cite{Toh01}, both lead to the inclusion of 2p2h configurations in their small amplitude limit thanks to collisional damping. 
Similarly, the RPA can be extended to include 2p2h excitations, leading to the so-called second-RPA (SRPA) \cite{dro90}. 
The additional residual interaction included in these approaches may affect the strength distributions of GRs as well as low-lying collective states \cite{gam10}. 

The improvement of computational power has recently led to realistic applications of the SRPA both with phenomenological interactions \cite{gam10,gam11,gam12} as well as forces derived from realistic two-body interactions \cite{pap09}. 
A common feature of these calculations is that the monopole strength is shifted toward lower energies by several MeV as compared to standard RPA calculations. 
As a result, the energy of the giant monopole resonance (GMR) is found at lower energies than the experimental values.
When realistic interactions are used \cite{pap09}, it could indicate a possible role of missing three-body interactions.
When effective interactions such as the Skyrme \cite{gam10,gam11} or Gogny \cite{gam12} interactions are used, this energy shift is shown to be strongly affected by large proton-neutron matrix elements of the residual interaction of the three-hole-one-particle (3h1p) type, coupling 1p1h with 2p2h configurations \cite{gam12}. 
Unfortunately, these terms of the residual interaction are not constrained in conventional fitting procedure of the effective interaction parameters. 
New parametrizations should then be developed to improve the predictive power of these beyond mean-field calculations. 
It should be noted also that, in principle, TDHF calculations in the non-linear regime are sensitive to 3h1p and 3p1h matrix elements of the residual interaction \cite{Sim03,Sim09}. 
However, their effect on, e.g., the spreading width of GRs has not been computed yet from TDHF evolutions.
Such calculations are beyond the scope of this work where we consider only 1p1h contributions to the GRs in order to investigate their direct decay in a simple and intuitive approach. 

Giant resonances lie usually at energies above the light particle emission threshold. Different kinds of decay are 
commonly distinguished. For instance, direct decay occurs when one particle is emitted leaving the daughter nucleus 
in a single hole state. However, the decay is sequential when two-particle two-hole (2p2h) configurations are populated 
by the residual interaction. The direct decay by nucleon emission is of particular interest as the emitted nucleon can 
give a glimpse on the microscopic structure of the GR. As such, it is an experimental tool of choice to study GRs~(see, 
e.g., Refs.~\cite{Har01,Str00,Zeg03,Hun07}).

Although a full description of the decay of GRs would involve beyond mean-field calculations to properly investigate the competition between direct and sequential decay, a qualitative description of the direct decay can be obtained with independent (quasi)particle approaches. 
In fact, if the collective motion is well described in the harmonic picture, i.e., by the (Q)RPA, then one can hope that such approaches would provide a good estimate of the direct decay contribution. 

The escape width can be computed at the independent (quasi)particle level if a proper treatment of the continuum is accounted for, i.e., in the continuum-(Q)RPA \cite{kre74,liu76,kam98,mat01,hag01,Kha02}. 
GR direct decay can also be investigated within the TDHF framework using large spatial grids \cite{Cho87,Pac88}.
In these calculations, only an estimate of $\Gamma^\uparrow$ and $\Gamma^L$ can be obtained assuming that the residual interaction coupling 1p1h states to 2p2h ones is weak for the associated modes. 
This approximation fails in heavier nuclei as the spreading width is known to dominate the total width in the GR region.
This is the case, for instance, in the chain of tin isotopes where $\Gamma^\downarrow$ accounts for almost half the total width of the GMR. 
For these nuclei, the independent (quasi)particle approach can only be used for qualitative studies. 
For lighter nuclei such as $^{16}$O, however, comparison between SRPA and RPA calculations show that, although the position of the GMR peak is sensitive to the residual interaction, the total width is almost unaffected by the addition of 2p2h contributions \cite{dro90,gam10,gam11,gam12,pap09}. 
It is possible, then, that in a first approximation, the direct decay of the GMR in $^{16}$O can be described with continuum RPA or TDHF calculations. 

The goal of the present work is to investigate the link between the 1p1h structure of giant resonances and their direct-decay 
pattern. The scalar-isoscalar Giant Monopole Resonance (GMR) is studied in $^{16}$O, $^{100}$Sn and $^{132}$Sn. 
These nuclei are closed-shell and, then, pairing correlations can be neglected. 
The response to a scalar monopole excitation in the small amplitude regime is computed using the time-dependent Hartree-Fock (TDHF) formalism \cite{Dir30}, based on a modern Skyrme functional~\cite{Sky56}. 
To study GMR direct-decay, the calculations are performed in a large spherical grid, allowing for a spatial separation of the nucleus from the emitted nucleons to the continuum. 
Similar TDHF calculations have been performed in the past with schematic energy density functionals to study GMR widths and lifetimes from direct decay~\cite{Cho87,Pac88}. 
These quantities were extracted from the time evolution of the number of emitted nucleons. 
In the present paper, new informations are obtained by computing energy spectra of the nucleons which are emitted in the direct-decay of the GMR. 
The structure of these spectra is then analyzed in terms of single-hole configurations of the $A-1$ daughter nucleus.

A detailed analysis is presented in the $^{16}$O case, for which the spreading width might be neglected, as discussed above. 
A comparison with the recently measured monopole strength function in this nucleus in the region 10-40~MeV \cite{lui01} will be presented.
The $^{100,132}$Sn cases are studied to infer the qualitative behavior of the direct decay as function of isospin.
In particular, we show that the GMR direct decay occurs only by proton (respectively neutron) emission in $^{100}$Sn, (resp. $^{132}$Sn), whereas both proton and neutron species contribute to the collective monopole vibration.

Theoretical framework and numerical details are presented in section~\ref{sec:framework}.
Results of the calculations are discussed in section~\ref{sec:results} 
before concluding in section~\ref{sec:conclusion}.

\section{Theoretical framework and numerical details}
\label{sec:framework}
\subsection{Time-dependent Hartree-Fock approach for collective vibrations}
\label{subsec:tdhf}

The TDHF equation reads~\cite{Rin80}
\begin{equation}
i \hbar \frac{d}{dt} \rho = \left[h\left[\rho\right],\rho\right],
\label{eq:tdhf}
\end{equation}
where $\rho$ is the one-body density matrix and $h\left[\rho\right]$ is the single-particle hamiltonian derived from the energy 
density functional $E[\rho]$ using
\begin{equation}
h_{ij}\left[\rho\right]=\frac{\delta E\left[\rho\right]}{\delta \rho_{ji}}.
\label{eq:HFhamiltonian}
\end{equation}
In the TDHF theory, the many-body state is constrained to be an independent particle state (e.g., a Slater determinant of the 
occupied single-particle states) which implies $\rho^2=\rho$ at all time. 

The one-body density matrix is used to compute expectation value of one-body operators for which TDHF is optimized~\cite{bal81}. 
Information on vibrational modes such as energy  can then be obtained through the evolution of their corresponding multipole moment. 
For instance, the excitation of a monopole mode $\ket{\nu}$ on top of the ground state induces an oscillation of the expectation value 
of the monopole moment~\cite{Rin80,Bla86}
\begin{equation}
\Fop=\sum_{\sigma,q}\int \mathbf{r}^2 \crea_{\mathbf{r}\sigma{q}} \anni_{\mathbf{r}\sigma{q}} d\mathbf{r}
\label{eq:operatorF}
\end{equation}
where $\crea_{\mathbf{r}\sigma{q}}$ ($\anni_{\mathbf{r}\sigma{q}}$) creates (annihilates) a nucleon in $\mathbf{r}$ with spin 
$\sigma$ and isospin $q$. In this mode, $\langle{\hat{F}}\rangle(t)$ oscillates at a frequency 
$\omega_\nu=(E_\nu-E_0)/\hbar$, where $E_0$ is the ground state energy and $E_\nu$ the energy of the excited state. 

The linearization of the TDHF equation leads to the RPA~\cite{Rin80}.
As a consequence, TDHF calculations in the small amplitude limit contain all the RPA residual interaction which is responsible for the collectivity of giant resonances. 
Using TDHF codes ensures that the same EDF is used to determine both the HF ground-state and the dynamics thanks to the structure of the TDHF equation. 
It is then equivalent to fully self-consistent RPA codes. 

In addition to energies of GRs, TDHF has been used to study their strength distribution~\cite{Blo79,Uma05,Mar05,Nak05,Alm05,Nak07,Ste07,Ste10}, 
their life-time and escape width~\cite{Cho87,Pac88}, and their anharmonicities~\cite{Sim03,Rei07,Sim09}. Escape width and Landau 
damping are accounted for in TDHF, however, two-body correlations responsible for the spreading width are neglected. Inclusion of 
pairing correlations is possible in the TDHF-Bogoliubov theory~\cite{Rin80,Bla86}, however its numerical applications~\cite{Ave08,Eba10,ste11} 
are time-consuming and prevent the use of large spatial grid as in the present study.

\subsection{Linear response theory\label{sec:linear}}

The nucleus is assumed to be in its ground state at initial time. An external field $\hat{V}_{{ext}}=\epsilon\hbar\eta(t)\Fop$ 
is applied, where $\epsilon$ is the intensity of the perturbation in fm$^{-2}$ and $\eta(t)$ is its time profile. We choose a delta-function 
for the latter: $\eta(t)=\delta(t)$. 
To the first order in $\epsilon$, the variation of the many-body state obtained from TDHF reads
\begin{equation}
|\delta\Psi(t)\> \simeq -i\epsilon \sum_\nu F_\nu e^{-iE_\nu t/\hb}  |\nu\> ,
\label{eq:1}
\end{equation}
where $|\nu\>$ are the eigenmodes and $F_\nu=\<\nu|\hat{F}|HF\>$ the transition amplitudes.

The evolution of an observable $\delta Q(t)=\expec{\Qop}(t)-\expec{\Qop}_0$ in the linear 
regime, where $\expec{\Qop}_0$ is the expectation value of $\Qop$ on the ground state evaluated at the Hartree-Fock level, can be written 
\begin{equation}
\delta Q(t) = -i\epsilon \sum_{\nu} Q_\nu^* F_\nu e^{-i \omega_\nu t} +c.c.,
\label{eq:Oop_evol}
\end{equation}
where $c.c.$ stands for {\it complex conjugated}. 
To analyze the spectral response of $\delta{Q}$, we introduce the quantity
\begin{equation}
R_{Q}(\omega) = \frac{-1}{\pi\epsilon}\int_{0}^{\infty}\delta Q(t)\sin({\omega t})dt.     \label{eq:FT_respfunc}
\end{equation}
which, in the particular case where $\Qop=\Fop$, reduces to the strength function
\begin{eqnarray}
R_{F}(\omega) &=& \sum_{\nu}\left|F_\nu\right|^2 \delta\left(\omega-\omega_\nu\right). \label{eq:Strength}
\end{eqnarray}

\subsection{Monopole vibrations in spherical symmetry}

The monopole moment of Eq.~(\ref{eq:operatorF}) is used in the external field. Assuming spherical symmetry of the ground state, 
only monopole modes are excited. It is then convenient to decompose the response of the nucleus in terms of the usual quantum 
numbers $l$ and $j$ using the operators
\begin{equation}
\Fop_{lj}=\sum_{n,n',m_j}\bra{nljm_j}\Fop\ket{n'ljm_j}\crea_{nljm_j} \anni_{n'ljm_j},
\label{eq:operatorFqlj}
\end{equation}
where the isospin is omitted to simplify the notation. The spectral response, $R_{F_{lj}}(\omega)$, defined in Eq.~(\ref{eq:FT_respfunc}),
can then be used to analyze microscopically the strength distribution of the GMR.

\subsection{Energy spectra of emitted nucleons}
\label{subsec:env}
\subsubsection{Calculation of the spectra}
After the application of the external perturbation $\epsilon\hbar\eta(t)\Fop$, the system deexcites by emission of nucleons
that escape from the inner nucleus region and propagate freely in the continuum\footnote{For now, we only focus on
neutrons that are not subject to the Coulomb field of the nucleus.}.
We are interested in exploring the spectra of emitted nucleons and their information content in terms of their parent excitation. \\
We assume that at a sufficiently large time $T$ after the initial external perturbation, once the excitations above the nucleon emission
threshold have decayed, any initially occupied time-dependent single particle wave function $\ket{\varphi_h(t\ge T)}$ can be spatially 
decomposed in two non-overlaping components: the first one is bound and localized in the nucleus region ($r<R_0$) 
$\ket{\varphi^{(in)}_h(t)}$, and the second one is an outgoing wave packet $\ket{\varphi^\uparrow_h(t)}$ localized in the outer region 
($r>R_0$), i.e.
\begin{equation}
\varphi_h^{\sigma q}(\mathbf{r};t\ge T) = \varphi^{\sigma q(in)}_h(\mathbf{r};t)+\varphi^{\sigma q\uparrow}_h(\mathbf{r};t),
\end{equation}
with $\varphi^{(in)}_h(\mathbf{r},r>R_0;t)=\varphi^\uparrow_h(\mathbf{r},r<R_0;t)=0$.
In the region $r>R_0$, nucleons evolve with the free propagator. Their (kinetic) energy spectra can thus be obtained by developping
these outgoing wave packets on the free wave basis, eigenstates of the free propagator with eigen-energy $\varepsilon_k=\frac{\hbar^2 k^2}{2m}$ 
with $m$ the nucleon mass and $k$ the modulus of the wave vector. In spherical coordinates, free waves can be written
\begin{eqnarray}
\bran{\mathbf{r}\sigma q}\ket{klj{m_j}q'}&=&\phi^\sigma_{klj{m_j}}\left(\mathbf{r}\right) \delta_{qq'}\\
                                     &=&\frac{1}{\mathcal{N}(k)} j_l(kr)\Omega^\sigma_{lj{m_j}}(\theta,\phi)\delta_{qq'},
\label{eq:spherical}
\end{eqnarray}
where $j_l(kr)$ is a spherical Bessel function of angular momentum $l$ and momentum $k$, $\Omega^\sigma_{lj{m_j}}(\theta,\phi)$
is a spherical spinor~\cite{Var88}, and $\mathcal{N}(k)$ is a normalization constant. 
As we consider scalar-isoscalar excitations [see Eq.~(\ref{eq:operatorF})] of spin saturated systems, we omit the spin and isospin in the following to simplify the notation. 
In addition, for GMR studies in spherical nuclei, spherical symmetry is conserved during the time evolution,
and time-dependent orbitals are still eigenvectors of the angular momenta operators. Overlaps are thus diagonal 
in $ljm_j$ and we will often omit them and their associated sum in the following as well.

An initially occupied single particle of quantum numbers $h\equiv nljm_j$
can thus be rewritten like
\begin{eqnarray}
\ket{\varphi^\uparrow_{h}(\mathbf{r};t)} = \int dk \tilde{\varphi}^\uparrow_{h}(k,t) \ket{k}
\end{eqnarray}
with
\begin{eqnarray}
\tilde{\varphi}^\uparrow_{h}(k,t)= \int \phi^*_{k}\left(\mathbf{r}\right) \varphi^\uparrow_{h}(\mathbf{r},t)d\mathbf{r}.
\label{eq:varphitilde}
\end{eqnarray}

The density of emitted nucleons per unit of $k$ at time $t>T$ with quantum numbers $l,j$ can thus be written 
\begin{eqnarray}
\tilde{\rho}^\uparrow_{l,j}\left(k,t>T\right) = \sum_{n,m_j} \int dk \tilde{\varphi}^{\uparrow*}_{nljm_j}(k,t) \tilde{\varphi}^\uparrow_{nljm_j}(k,t).
\end{eqnarray}
As emitted nucleons propagate freely, this quantity is time independent for $t>T$. 
The density per energy unit can then be obtained from $\rho(\varepsilon)d\varepsilon=\tilde{\rho}(k)dk$.
The energy spectrum of emitted nucleons (with quantum numbers $l$ and $j$) then reads 
\begin{eqnarray}
\rho^\uparrow_{l,j}(\varepsilon)=\tilde{\rho}^\uparrow_{l,j}(\sqrt{2m\varepsilon}/\hbar)\frac{dk}{d\varepsilon}.
\label{eq:EkDOS}
\end{eqnarray}
\subsubsection{Content of the spectra of emitted nucleons}
\label{sec:TDAmapping}
We now focus on the information content of these spectra of emitted nucleons.
In TDHF, the many-body state is constrained to be a Slater determinant at any time $t$:
\begin{equation}
|\Psi(t)\> = \( \prod_{i=1}^A \hat{a}_i^\dagger(t)\)|-\> ,\label{eq:Psi}
\end{equation}
where $\hat{a}_i(t)$ creates a particle in the state $|\varphi_i(t)\>$.  
According to the Thouless theorem \cite{Tho60}, such a state can be expressed as 
\begin{equation}
|\Psi(t)\> = \mathcal{N}(t) e^{\sum_{ph}Z_{ph}(t)\hat{a}_p^\dagger \hat{a}_h} |HF\>,
\end{equation}
where $\mathcal{N}(t)$ is a normalization constant, $\hat{a}_h$ annihilates a hole state, and $\hat{a}_p^\dagger$ creates a particle state.
 In the first order in $\epsilon$, the variation of $|\Psi(t)\>$ reads
\begin{equation}
|\delta\Psi(t)\> \simeq -i\epsilon e^{-iE_0t/\hb}\sum_{ph} Z'_{ph}(t)\hat{a}_p^\dagger\hat{a}_h |HF\>,\label{eq:delpsi}
\end{equation}
with $Z=-i\epsilon Z'$. 
From Eq.~(\ref{eq:1}), we get 
\begin{equation}
Z'_{ph} (t)= \sum_{\nu} e^{-i\omega_\nu t} F_{\nu} \<HF|\hat{a}_h^\dagger \hat{a}_p|\nu\> .\label{eq:Z'}
\end{equation}

Eq.~(\ref{eq:Psi}) can be written
\begin{equation}
|\Psi(t)\>=\left[ \prod_{h=1}^A \(e^{-\frac{i}{\hb}e_ht}\hat{a}_h^\dagger + \delta\hat{a}_h^\dagger(t)\)\right]|-\>, \label{eq:HF-basis}
\end{equation}
From Eqs.~(\ref{eq:delpsi}) and~(\ref{eq:Z'}), we see that, in the first order in~$\epsilon$, the variation of the single-particle states obey
\begin{eqnarray}
|\delta\varphi_h(t)\> &\simeq & -i\epsilon e^{-\frac{i}{\hb}e_ht} \sum_p Z'_{ph}(t) |p\> \nonumber \\
&=&  -i\epsilon \sum_\nu e^{-\frac{i}{\hb}(\hb\omega_\nu+e_h)t} F_\nu \sum_p \<HF|\hat{a}_h^\dagger \hat{a}_p|\nu\> |p\>. \nonumber \\ \label{eq:phi}
\end{eqnarray}

At large time $t>T$, the unbound part of $|\varphi_h(t)\>$ is a free wave packet which can be expressed as
\begin{eqnarray}
\ket{\varphi^\uparrow_{h}(t)} 
&\simeq& - i \epsilon \int dk f_h(k) e^{-\frac{i}{\hbar}\varepsilon_k t} \ket{k} , \label{outkdecomp}
\end{eqnarray}
where $\varepsilon_k=\hbar^2 k^2 / 2m$ is the free propagator that acts in the outer region
(no Hartree-Fock fields). 
It is then convenient to write Eq.~(\ref{eq:phi}) in the basis of $|k\>$ while keeping only the unbound part by taking the sum  over unbound states only, i.e., with $\hb\omega_\nu > S=\min\(S_p,S_n\)$, where $S_{p,n}$ are proton and neutron separation energies:
\begin{eqnarray}
\ket{\varphi^\uparrow_{h}(t)} 
&\simeq& - i \epsilon \sum_{\nu,\hb\omega_\nu>S} e^{-\frac{i}{\hbar} \left(\hbar\omega_\nu +e_h\right)t}F_\nu \int dk \chi^\nu_{h}(k) \ket{k},\nonumber \\
\label{outkdecompTDA}
\end{eqnarray}
with $\chi_h^\nu(k) = \sum_p \<k|p\>\<HF|\hat{a}_h^\dagger\hat{a}_p|\nu\>$.

Identifying Eqs.~(\ref{outkdecomp}) and (\ref{outkdecompTDA}), we have
\begin{eqnarray}
f_{h}(k) = \sum_{\nu,\hb\omega_\nu>S} e^{-\frac{i}{\hbar} \left(\hbar\omega_\nu +e_h-\varepsilon_k\right)t}F_\nu \chi^\nu_{h}(k).
\end{eqnarray}
To conserve energy, $f_h(k)$ must be constant for $t\ge T$. 
 This implies that
\begin{eqnarray}
\chi^\nu_{h}(k) \propto \delta(\hbar\omega_\nu +e_h-\varepsilon_k) \label{eq:khnuequivalence}
\end{eqnarray}
and
$f_h(k) = \sum_\nu F_\nu \chi^\nu_{h}(k).
$
As a result
%\footnote{We also notice that in this model, Eq.~(\ref{eq:khnuequivalence}) implies that $\varepsilon_k$ accounts for the residual interaction, i.e. it does not correspond simply to the energy of an Hartree-Fock particle state.}
we have
\begin{equation}
\ket{\varphi_h^{\uparrow}(t>T)} \simeq -i\epsilon \int dk \sum_{\nu,\hb\omega_\nu>S} F_\nu \chi^\nu_{h}(k) e^{-\frac{i}{\hb}\varepsilon_k t} 
\ket{k},
\end{equation}
and the  density of emitted nucleons per unit of $k$  reads, for $t>T$,
\begin{eqnarray}
\tilde{\rho}^\uparrow(k) 
&\simeq&\epsilon^2 \sum_{h=1}^A\sum_{\nu,\hb\omega_\nu>S} \left|F_\nu\right|^2 \int dk' \left|\chi^\nu_{h}(k')\right|^2 \delta(k-k').\nonumber\\
\label{eq:rhokTDA}
\end{eqnarray}
Using Eq.~(\ref{eq:EkDOS}) and 
$$\delta\left(k-\sqrt{\frac{2m\varepsilon_{k'}}{\hbar^2}}\right)=\delta\left(\frac{\hbar^2k^2}{2m}-\varepsilon_{k'}\right)\frac{dE}{dk},$$
we get the density of emitted nucleons per unit of energy
\begin{eqnarray}
\rho^\uparrow(E) 
\simeq 
\epsilon^2 \sum_{h=1}^A\sum_{\nu,\hb\omega_\nu>S} \left|F_\nu\right|^2 \int dk \left|\chi^\nu_{h}(k)\right|^2 \delta(E-\varepsilon_{k}) \nonumber\\
%\delta(E-\hbar\omega_\nu-e_h)\nonumber\\
 \label{eq:trans1}
%=\epsilon^2 \sum_\nu \left|\bra{\nu}\Fop\ket{0}\right|^2 \int dk \left|X^\nu_{kh}\right|^2 \delta(E-\hbar\omega_\nu+e_h)\label{eq:trans2}
\end{eqnarray}
%where Eq.~(\ref{eq:trans1}) has been obtained using 
%$$\delta\left(k-\sqrt{\frac{2m\varepsilon_{k'}}{\hbar^2}}\right)=\delta\left(\frac{\hbar^2k^2}{2m}-\varepsilon_{k'}\right)\frac{dE}{dk}$$
%and Eq.~(\ref{eq:EkDOS}), and where we have used Eq.~(\ref{eq:khnuequivalence}) to obtain Eq.~(\ref{eq:trans2}).

Using Eq.~(\ref{eq:khnuequivalence}) and replacing $E\rightarrow E+e_h$,
we can define the shifted spectra by 
\begin{eqnarray}
 \rho^{\uparrow(s)}(E) &\simeq&
\epsilon^2 \sum_{\nu,\hb\omega_\nu>S} \left|F_\nu\right|^2 \delta\left(E-\hbar\omega_\nu\right)\sum_{h=1}^A \int dk \left|\chi^\nu_{h}(k)\right|^2\nonumber \\
&\simeq&
\epsilon^2 \sum_{\nu,\hb\omega_\nu>S} \left|F_\nu\right|^2 \delta\left(E-\hbar\omega_\nu\right),
\end{eqnarray}
where we have used the normalization condition for excited states.
We thus recover the expression of the strength function [Eq.~(\ref{eq:Strength})] up to the quadratic factor $\epsilon^2$.
This shows that within this approach, the emitted nucleons keep track of the information 
content of the decayed mode. \\
\indent Finally, we remind that the decomposition in free waves is well-adapted to obtain energy spectra of neutrons. 
Outside the nucleus region, protons are still subject to the Coulomb interaction with the remaining nucleus. In the numerical 
applications, we use Coulomb wave-functions instead of Bessel wave-functions in Eq.~(\ref{eq:spherical}) to determine proton 
spectra. 

\subsection{Numerical details}
\label{subsec:numerics}

The evolution of the occupied single-particle wave-function is determined from the TDHF equation in a spherical mesh of radius 
$R_{box}=600$~fm, with a radial spacing $\delta r=0.3$~fm and hard boundary conditions. The SLy4 parametrization~\cite{Cha98} of the Skyrme 
EDF~\cite{Sky56} is used. In a first step, the HF initial condition is computed in a $R_0=30$~fm box with the same radial 
spacing. At initial time, all single-particle wave-functions vanish for $r>R_0$. 

The spectrum of emitted nucleons is computed in the outer region from $R_0$ to $R_{box}$. 
The maximum evolution time is set to $T=2250$~fm/$c$. 
This choice of $T$ and $R_{box}$ ensures that particles with less than $E_{max}\simeq33$~MeV do not have time to reach the edge of the box. 
In the present applications, no emitted particle with more than 25~MeV were found. 
Consequently, the monopole response and the spectra of emitted nucleons are not affected by spurious reflection on the hard box boundaries~\cite{Rei06}.

Note that the upper limit of the integral in the definition of the monopole moment in Eq.~(\ref{eq:operatorF})
is chosen to be $R_{nucl}=4\,R_{rms}$ for $^{16}$O and $R_{nucl}=R_0$ for $^{100,132}$Sn to avoid a divergence 
of $\delta{F}$ due to emitted nucleons~\cite{Str79}.
%CSfinal<
The sensitivity of the results with this upper limit is discussed in Section~\ref{sec:results}.
%CSfinal>

To account for the finite size of the box, only spherical free waves (for neutrons) and Coulomb waves
(for protons) vanishing at $R_{box}$ are considered in Eq.~(\ref{eq:spherical}). The latter are computed using 
the direct integration technique of~\cite{Ben05}. Their normalization constant $\mathcal{N}$ is determined assuming 
that they vanish for $r\ge R_{box}$. As a consequence, the momentum of each partial wave with quantum number $j,l$ 
can take only discrete values. In order to simulate the density of states of free (or Coulomb) waves within the box, 
$\rho^{\uparrow}_{lj}$ is obtained by convoluting the discretized density of emitted nucleons with a normalized 
Gaussian distribution in energy with a standard deviation of $150$~keV.

\section{Results}
\label{sec:results}
\subsection{Detailed analysis of the GMR in $\Ox{16}$}
\begin{figure}[!h]
\begin{center}
\includegraphics{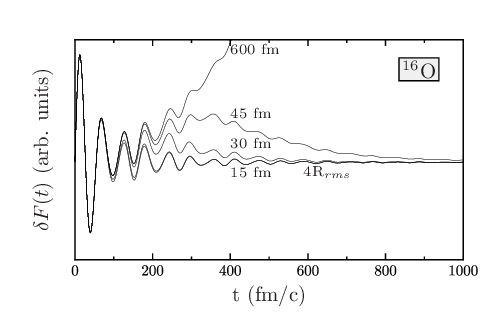}
\end{center}
\caption{Evolution of the monopole moment in $\Ox{16}$ after a monopole boost with different values of the upper limit $R_{cut}$ in the integral of Eq.~(\ref{eq:operatorF}).}
\label{fig:Rcut-T}
\end{figure}

\begin{figure}[!h]
\begin{center}
\includegraphics{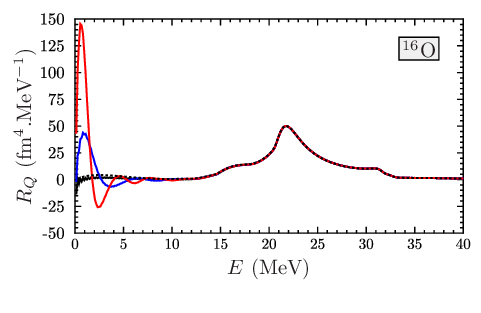}
\end{center}
\caption{\emph{(Color online)} Strength function of the GMR in $\Ox{16}$ computed from the time evolutions 
of the monopole moment $\delta{F}(t)$ with $R_{cut}=4R_{rms}$ (solid black line), 30~fm (blue line), and 45~fm (red line).
The dashed black line has been obtained with $R_{cut}=4R_{rms}$ and a boost velocity $2\varepsilon$.}
\label{fig:Rcut-E}
\end{figure}

To illustrate the method, let us start with a microscopic analysis of the GMR direct decay in $\Ox{16}$. The 
evolution of the expectation value $\delta{F}(t)=\langle{F}\rangle(t)-\langle{F}\rangle_0$ of the observable 
$\Fop$ given in Eq.~(\ref{eq:operatorF}), where the integral is performed up to %CSfinal< 
several distances $R_{cut}$,
%$4\,R_{rms}$
after a monopole 
excitation in linear regime is shown in Fig.~\ref{fig:Rcut-T}. 
The variation of the average value of the  monopole moment for large $R_{cut}$ is due to emitted particles. 
When the latter evolve in the $r>R_{cut}$ region, they do not contribute to the monopole moment anymore and the $\delta F(t)$ converge at large times (except for $R_{cut}=600$~fm which is at the edge of the box). 
%Fig.~\ref{fig:O16_evol}. 
The strength functions associated to $\delta{F}(t)$ and obtained from Eq.~(\ref{eq:FT_respfunc}) with $Q=F$ are plotted 
in Fig.~\ref{fig:Rcut-E} for $R_{cut}=4R_{rms}$, 30~fm, and 45~fm.
We see that they all converge in the energy range of interest for the GMR, i.e., $E>10$~MeV. 
The following analysis for $^{16}$O is performed with $R_{cut}=4R_{rms}$.
A strength function obtained with a double intensity excitation $2\varepsilon$ is also shown in Fig.~\ref{fig:Rcut-E}  (dashed line). 
It is identical to the result with a boost velocity $\varepsilon$, indicating that the calculations  are performed in the linear regime. 
%CSfinal>

\begin{figure}[!h]
\begin{center}
\includegraphics{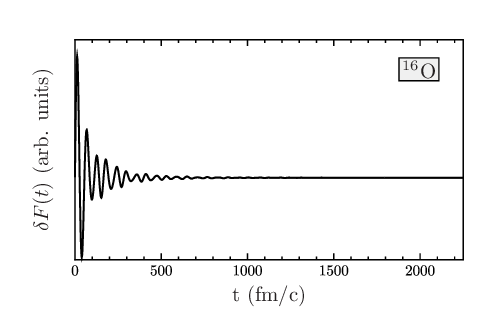}
\end{center}
\caption{Evolution of the monopole moment in $\Ox{16}$ after a monopole boost in linear regime with $R_{cut}=4R_{rms}$ and with the SLy4 parametrization of the Skyrme EDF.}
\label{fig:O16_evol}
\end{figure}

The evolution of  $\delta{F}(t)$ over the maximum evolution time is shown in Fig.~\ref{fig:O16_evol}. 
A damped oscillation associated to the GMR is observed, with an average period $T_{GMR}\simeq57.9$~fm/$c$ 
%%%%%%%%%%%%%%%%%%%%%%%%%%%%%%%%%%%%%%%%%%%%%%%%%%%%%%%%%%%%%%%%%%%%%%%%%%%%%%%%%%%%%%%%%%%%%%%%%%%%%%
% NOTE : This has been computed by taking the maxima of the [0:950] t-segment, giving T=57.9368 fm/c %
%%%%%%%%%%%%%%%%%%%%%%%%%%%%%%%%%%%%%%%%%%%%%%%%%%%%%%%%%%%%%%%%%%%%%%%%%%%%%%%%%%%%%%%%%%%%%%%%%%%%%%
corresponding to an energy $E_{GMR} = 2\pi\hbar/T_{GMR}$ $\simeq 21.4$ MeV.
In a coherent state picture (see, e.g., \cite{sim01}), the number of excited phonons is proportional 
to the square of the oscillation amplitude. The damping of the oscillation is then a signature for GMR decay. 
Here, the decay occurs by nucleon emission as $E_{GMR}$ is greater than the proton and neutron separation 
thresholds in $^{16}$O. The description of nucleon emission is possible if the continuum is properly 
accounted for. This is ensured, here, by the large size of the spherical mesh to prevent spurious effects 
coming from reflected flux on the edge of the box~\cite{Rei06}.

\begin{figure}[!h]
\begin{center}
\includegraphics{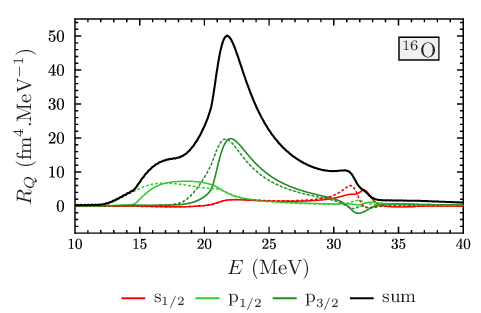}
\end{center}
\caption{\emph{(Color online)} Strength function of the GMR in $\Ox{16}$ (black solid line) computed from the time evolution 
of the monopole moment $\delta{F}(t)$ shown in Fig.~\ref{fig:O16_evol} with the SLy4 parametrization of the Skyrme EDF. The spectral responses $R_{F_{lj}}$ associated to the single-particle quantum 
numbers $l$ and $j$ and labelled by their spectroscopic notation are plotted in colored solid (dashed) lines for neutrons 
(protons).}
\label{fig:S0_O16}
\end{figure}

The strength function associated to $\delta{F}(t)$ is plotted 
in Fig.~\ref{fig:S0_O16} (black solid line). The main peak at $\sim{E}_{GMR}$ is surrounded by two shoulders 
at $\sim17$ and $\sim31$~MeV. In order to get a deeper insight into the microscopic origin of these structures, the 
spectral responses of the $\delta{F}_{lj}(t)$ [see Eq.~(\ref{eq:operatorFqlj})] following the same monopole boost are 
computed from Eq.~(\ref{eq:FT_respfunc}) and plotted with colored lines in Fig.~\ref{fig:S0_O16}. We have checked 
numerically that the sum of these spectral functions is equal to the total strength function $R_F$, as expected. This 
decomposition shows that the low-energy shoulder and the main peak are mainly due to  $\ljp{1}$ and $\ljp{3}$ orbitals, 
respectively.
The structure of the high-energy shoulder is more complicated. It involves a constructive contribution of the $\ljs$
orbitals. The dominant role of the  $\ljs$ neutron and proton contributions in the high-energy shoulder has already 
been noted using similar techniques but other EDF parametrizations~\cite{Pac88,Ste10}.
In addition, we see that the $\ljp{3}$ orbitals reduce the strength at $\sim32$~MeV because $R_{F_{p3/2}}$ is negative in 
this region. Unlike strength functions ($\Qop=\Fop$ case in section~\ref{sec:linear}), spectral responses of $\Qop\ne\Fop$ 
are not necessarily strictly positive. In particular, it happens at frequencies $\omega\simeq\omega_\nu$ when 
$Re(\langle0|\hat{Q}|\nu\rangle\langle\nu|\hat{F}|0\rangle)<0$. In this case, the $\omega$-component of $F(t)$ and $Q(t)$ 
oscillate with opposite phases. For the monopole response in $^{16}$O, this means that the  contribution of the $\ljp{3}$ 
orbitals at $\sim32$~MeV ''fight against'' the mean-field which oscillates like $\delta{F}(t)$, generating a destructive 
contribution to the monopole strength function.

\begin{figure}[!h]
\begin{center}
\includegraphics[width=8cm]{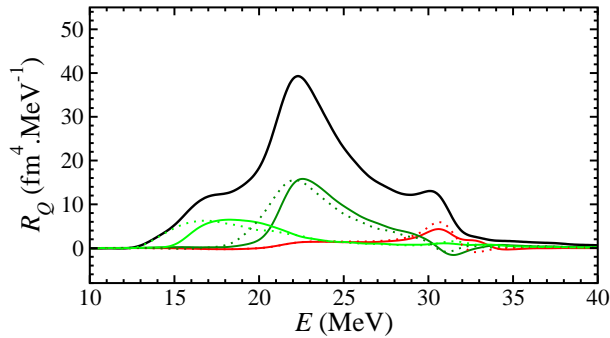}
\end{center}
\caption{\emph{(Color online)} Same as Fig.~\ref{fig:S0_O16} with the SGII parametrization of the Skyrme EDF.}
\label{fig:SGII}
\end{figure}

%CSfinal<
This microscopic decomposition of the strength function may vary with the parametrization of the Skyrme EDF. 
Indeed, the latter is usually fitted on global properties of nuclear systems, while the single-particle levels are not directly constrained. 
It is then worth comparing the above results with another parametrization of the Skyrme EDF. 
We chose the SGII parametrization which has been introduced to study compression modes in nuclei with a compression modulus $K_\infty\simeq 217$ MeV in infinite nuclear matter \cite{SGII}. 
This value is slightly higher with SLy4, i.e., $K_\infty\simeq 230$~MeV. 
The results, shown in Fig.~\ref{fig:SGII}, are very similar to the calculations with the SLy4 parametrization. 
%CSfinal>

\begin{figure}
\begin{center}
\includegraphics[width=8cm]{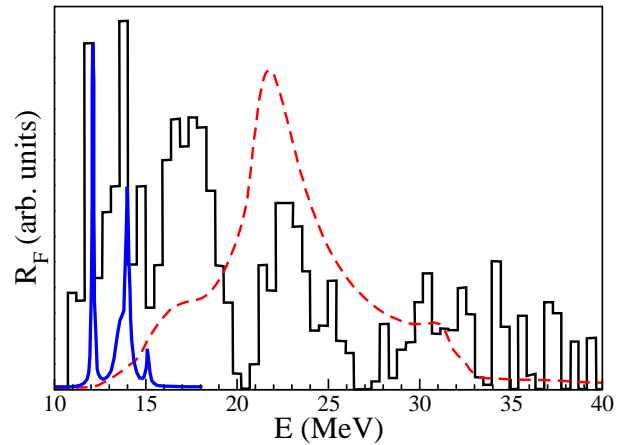}
\end{center}
\caption{\emph{(Color online)} Experimental monopole strength distribution in $\Ox{16}$  from Ref.~\cite{lui01} (black solid line). 
The TDHF strength function is reported in dashed red line. The normalization is arbitrary. The thick solid blue line shows monopole strength expected from states with $\alpha$ clustering (From Ref.~\cite{yam12}).}
\label{fig:GMR}
\end{figure}

The monopole strength in $^{16}$O has been measured recently between 10 and 40 MeV energy excitation \cite{lui01}. 
The results are reported in Fig.~\ref{fig:GMR} (black solid line) and compared to the TDHF monopole strength distribution (red dashed line). 
Approximatively half of the energy weighted sum rule has been found in this energy region. 
The rest of the strength distribution is expected to be found in low lying states at $E<10$~MeV, which could not be measured with the experimental setup, and at energies $E>40$~MeV. 
Note that the low-lying $0^+$ states, such as the one at $\sim6$~MeV in $^{16}$O, can be reproduced only if 2p2h contributions are included in the model ~\cite{gam10}. 
As a result, they are not seen in the present TDHF calculations. 
%Note also that large error bars (not shown in the figures) were obtained in the high energy part ($E>30$~MeV)~\cite{lui01}. 

The present calculations do not reproduce the peaks in the $10<E<15$~MeV region.
In fact, the latter have been recently interpreted in terms of $\alpha$ cluster states (see thick solid blue line in Fig.~\ref{fig:GMR}) \cite{yam12} which are outside the space of 1p1h states. 
Note that monopole vibrations of $\alpha$ cluster configurations in light nuclei have also been recently investigated with Fermionic Molecular Dynamics (FMD) calculations \cite{fur10}. 

The strength distribution of the GMR, as predicted by the present TDHF calculations, extends from $\sim15$ to $33$~MeV.  
As shown in Fig.~\ref{fig:GMR}, the experimental strength in this region clearly exhibits three peaks at positions which could tentatively be associated to the $p_{1/2}$ ($E\sim15-20$~MeV), $p_{3/2}$ ($E\sim20-26$~MeV), and $s_{1/2}$ ($E\sim28-33$~MeV) contributions of the TDHF response in Fig.~\ref{fig:S0_O16}. 
Similar three peak structures in the $^{16}$O GMR have been obtained in several microscopic calculations \cite{ma97,gam10,gam11,gam12,pap09}. 
The relative weight of the first peak is  underestimated by the TDHF calculations.
A possible explanation is that states with 2p2h contributions might be present, and even dominate, in this region of the spectrum. 

It is also interesting to note that the width of these peaks (in particular the one in the $20-26$ MeV region) seems to be of the same order than the TDHF prediction. 
Assuming that the present TDHF calculations provide a good estimate of $\Gamma^\uparrow$ and $\Gamma^L$, this would imply that $\Gamma^\downarrow$ is only a correction to the total width of the GMR in $^{16}$O.
Of course, $\Gamma^\downarrow$ could not be neglected in heavier nuclei such as tin isotopes where it account for about half of the total width \cite{tse09}. 

\begin{figure}[!h]
\begin{center}
\includegraphics{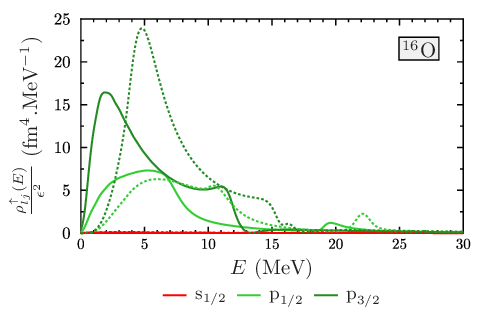}
\end{center}
\caption{\emph{(Color online)} Neutron (solid lines) and proton (dashed lines) direct-decay spectra of the GMR in $\Ox{16}$ for 
each set of single-particle quantum numbers $l$ and $j$, given in spectroscopic notation.}
\label{fig:spec_decay_O16}
\end{figure}

The next step of our analysis is to study the decay properties of the GMR and to investigate their relationship with 
the previous microscopic decomposition. The spectra of emitted nucleons are computed at $T=2250$~fm/$c$ after the boost, e.g., 
when the monopole oscillation is fully damped (see Fig.~\ref{fig:O16_evol}). We use the method described in 
Sec.~\ref{subsec:env}. The proton and neutron spectra for each set of $l$ and $j$ are plotted in 
Fig.~\ref{fig:spec_decay_O16}. We have checked numerically that, in the linear regime, these spectra are 
quadratic in the intensity of the boost~$\epsilon$~\cite{Cho87}.

First we note that proton and neutron spectra have similar global features, as expected for light $N=Z$ nuclei, 
although the proton spectra are shifted by $\sim3$~MeV due to the Coulomb repulsion.
In addition, the Coulomb barrier, which is $B_c\simeq1.7$~MeV in $^{16}$O, reduces strongly the emission of 
protons at energy $E<B_c$ due to  the exponentially decreasing tunneling probability. We also note that there 
is no relevant emission of $\ljs$ orbitals. In fact, the single-particle energy of the $1s_{1/2}$ orbitals, 
which are the only occupied $\ljs$ single-particle states in the HF ground-sate of $^{16}$O, are $-32.4$~MeV 
for protons and $-36.2$ for neutrons (see table~\ref{tab:16Oesp}). As a result, the 1p1h states with a hole in 
a $1s_{1/2}$ orbital, and contributing to the GMR microscopic structure, are bound for neutrons and below $B_c$ for protons. Indeed, 
the different structures in the GMR spectrum (see Fig.~\ref{fig:S0_O16}) are located at energies below $\sim33$~MeV, 
which is not sufficient to emit a $1s_{1/2}$ single-particle.

\begin{table}
\caption{\label{tab:16Oesp} Energies (in MeV) of the occupied single-particle states in the HF ground state of $^{16}$O with the SLy4 parametrization.}
\begin{center}
\begin{tabular}{ccc}
\noalign{\smallskip}\hline
 s.p. state & proton & neutron\\
\hline\noalign{\smallskip}
$1s_{1/2}$ &-32.4& -36.2\\
$1p_{3/2}$ &-17.1& -20.6\\
$1p_{1/2}$ &-11.2& -14.5\\
\hline\noalign{\smallskip}
\end{tabular}
\end{center}
\end{table}

\begin{figure}[!h]
\begin{center}
\includegraphics{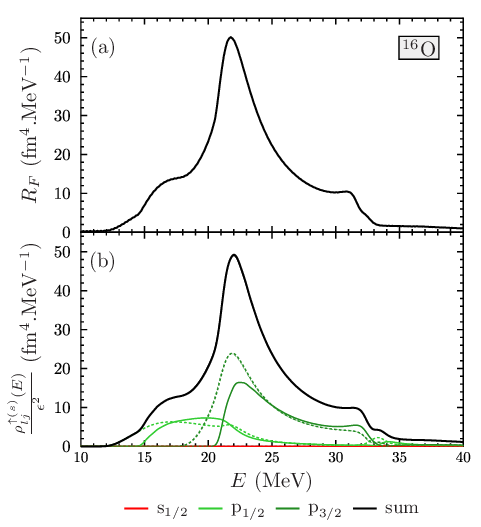}
\end{center}
\caption{\emph{(Color online)} (a) Strength function of the GMR in $^{16}$O obtained from the time evolution of the monopole moment. 
(b) Spectra of emitted neutrons (colored solid lines) and protons (colored dashed lines) ''shifted'' by the energy of the initially 
occupied single-particle state (see text). Their sum is shown in black solid line.}
\label{fig:S0_decay_O16}
\end{figure}

Let us now attempt to reconstruct the GMR strength function from the spectra of emitted nucleons shown in 
Fig.~\ref{fig:spec_decay_O16}. In a pure harmonic picture, the GMR is a coherent sum of 1p1h states with a 
difference $\sim{E}_{GMR}$ between the particle and the hole energies. 
However, nucleons emitted by direct decay have only an energy $\sim$ to the one of a particle-state, which 
is obviously smaller than $E_{GMR}$ because the energy of the hole-state is negative. 
To approximately reconstruct the GMR strength function, it is then necessary to shift the energy of the emitted 
nucleons by the energy of their associated hole-state, that is, the energy of the initially occupied single-particle 
state. The latter are given in table~\ref{tab:16Oesp} for $^{16}$O at the HF level.
The ``shifted'' spectra of emitted nucleons for each set of $l$ and $j$, noted $\rho^{\uparrow(s)}_{lj}$, 
obtained from this procedure, are plotted in Fig.~\ref{fig:S0_decay_O16}-b, together with their sum.
It is striking to see that, not only the shape, but also the magnitude of the latter is very similar to the 
strength function shown in Fig.~\ref{fig:S0_O16} and recalled in Fig.~\ref{fig:S0_decay_O16}-a, although the 
two spectra have been obtained from totally different quantities, i.e., the spectra of emitted nucleons and 
the time evolution of the monopole moment.
A theoretical justification of these similarities is outlined in section~\ref{sec:TDAmapping}.

Similarly to the microscopic decomposition of the strength function obtained from the time evolution of the 
monopole moment (see Fig.~\ref{fig:S0_O16}), we see in Fig.~\ref{fig:S0_decay_O16}-b that the low energy 
shoulder and the main peak are associated to $\ljp{1}$ and $\ljp{3}$ orbitals, respectively. Again, the 
situation for the high energy shoulder is more complicated. Although it is mainly due to $\ljs$ orbitals 
in the monopole response (see Fig.~\ref{fig:S0_O16}), it is in fact associated to the emission of $p$-orbitals 
in Fig.~\ref{fig:S0_decay_O16}-b. As discussed before, this is because the energy of the GMR is not sufficient 
to promote a $1s_{1/2}$ single-particle to the continuum. However, the RPA residual interaction couples the 
responses associated to the different single-particle quantum numbers to produce a collective monopole vibration. 
As a result, the vibration of $s$-orbitals can be coupled to $p$-unbound-states contributing to the direct decay. 
Such coupling might be the reason for the complex competition between constructive and destructive contributions 
in the high-energy shoulder observed in Fig.~\ref{fig:S0_O16}.

\subsection{$A=100$ and $132$ tin isotopes}

\begin{figure}[!h]
\includegraphics{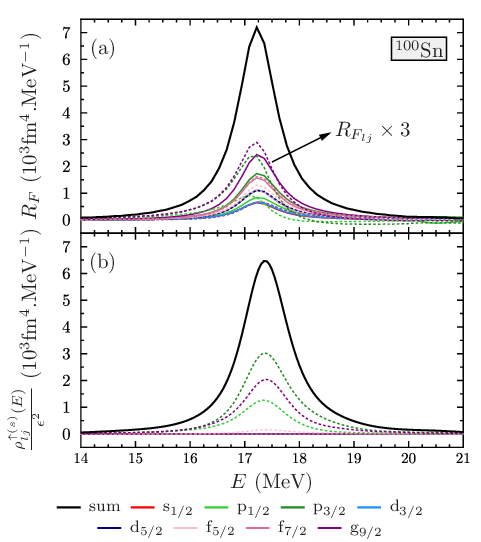}
\caption{\emph{(Color online)} (a) Strength function $R_F$ (black solid lines) of $\Sn{100}$. Its single-particle 
decomposition, multiplied by a factor of three ($R_{F_{lj}}\times 3$), is represented with colored solid (dashed) 
lines for neutrons (protons). (b) Spectra of emitted particles ''shifted'' from their single-particle energies 
are plotted in colored solid (dashed) lines for neutrons (protons). Their sum appears in black.}
\label{fig:Tin100}
\end{figure}

\begin{figure}[!h]
\includegraphics{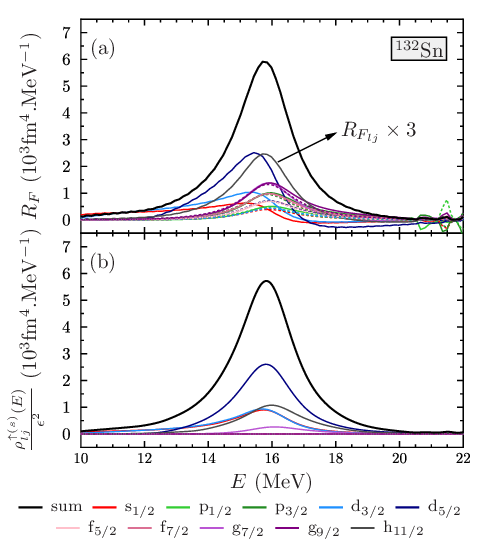}
\caption{\emph{(Color online)} Same as Fig.~\ref{fig:Tin100} for $\Sn{132}$.}
\label{fig:Tin132}
\end{figure}

Similar analyses have been performed for the GMR in  $^{100,132}$Sn nuclei. The strength functions obtained from the time 
evolution of the monopole moment are plotted with black solid lines in the upper panels of Fig.~\ref{fig:Tin100} 
and \ref{fig:Tin132}. The reconstructed GMR strength, obtained from the emission spectra are plotted with black solid lines 
in the lower panels. As in the $^{16}$O case, the agreement between the two methods is excellent.

The gross feature of the strength is similar for the two isotopes, with a single peak centered at $E_{GMR}=17.2$~MeV for 
$\Sn{100}$ and $15.8$~MeV for $\Sn{132}$ (see table~\ref{tab:Sn}), although their widths vary more importantly. It is 
striking, however, that the decompositions in terms of single-particle quantum numbers exhibit some important differences 
between the two techniques. Although both proton and neutron single-particle orbitals participate to the vibration (see 
upper panels in Fig.~\ref{fig:Tin100} and \ref{fig:Tin132}), the decay only occurs by neutron emission for $\Sn{132}$, 
whereas $\Sn{100}$  decays through proton emission only. This can be understood by the fact that the proton (neutron) separation 
energy increases (decreases) with the number of neutrons. These quantities are given in table~\ref{tab:Sn}, together with the 
Coulomb barrier for protons and the position of the peak energy associated to the GMR energy. Indeed, the binding energy of 
neutrons in $^{100}$Sn is of the order of $E_{GMR}$ and proton emission is favored, while the combined effects of an increasing $S_p$ 
and the Coulomb barrier $B_c$ prevent proton emission in $\Sn{132}$. 

\begin{table}
\caption{\label{tab:Sn} Neutron and proton separation energies and Coulomb barriers in Sn isotopes.
The GMR energy corresponds to the position of the peak in the strength function. All energies are in MeV.}
\begin{center}
\begin{tabular}{ccccccc}
\hline\noalign{\smallskip}
  & $S_p$ & $B_{c}$ & $S_p+B_{c}$ & $S_n$ & $E_{GMR}$\\
\hline\noalign{\smallskip}
$^{100}$Sn &  3.1 & 8.1 & 11.4 & 16.9 & 17.2 \\
$^{132}$Sn & 15.6 & 7.4 & 23.0 &  7.7 & 15.8 \\
\noalign{\smallskip}\hline
\end{tabular}
\end{center}
\end{table}

As in the case of the high-energy shoulder in the $^{16}$O spectra of figures~\ref{fig:S0_O16} and~\ref{fig:S0_decay_O16}, 
the difference in the microscopic decomposition of the GMR strength with the two techniques is a signature of the RPA residual 
interaction which couples the bound particle-hole states to the unbound ones, allowing their decay. In particular, the vibration 
of the protons (neutrons) in $^{132}$Sn ($^{100}$Sn) decays via neutron (proton) emission thanks to the collective vibration 
of both proton and neutron mean-fields.


\section{Conclusions}
\label{sec:conclusion}
Using a simple single-particle decomposition, the microscopic structure and (direct) decay of the GMR in some spherical nuclei 
has been studied in the harmonic picture using TDHF calculations. 
The GMR strength is usually obtained, in TDHF, using the Fourier analysis of the real-time 
small amplitude monopole response.
We showed that, assuming that 2p2h and higher order configurations do not contribute significantly to the strength function, it can be also approximately reconstructed by shifting the energy spectra of emitted 
nucleons by their corresponding initial hole single-particle energy. 

On this ground, we studied the microscopic structure of the GMR strengths obtained by these two techniques (standard real-time 
response and reconstruction from emitted nucleons), by decomposing it onto the single-particle quantum numbers $l$ and $j$. 
Although both techniques lead to almost identical monopole strengths, their microscopic structure may be very different. This 
is understood by the fact that, although GMR energies are usually above the nucleon emission thresholds, only part of the 
particle-hole states lie in the continuum. 
Those which are bound are coupled to unbound states via the RPA residual interaction which is responsible for the collectivity 
of the vibration. This is illustrated, e.g., in monopole spectra of neutron-rich tin isotopes: although both protons and neutrons 
contribute to the monopole vibration, only neutrons can be emitted because protons are more bound and have to overcome the Coulomb 
barrier. 

A major improvement of the method would be to include two-body correlations, like pairing interaction and collision terms.  
Pairing correlations are  expected to affect particle decay as compared to the independent particle picture~\cite{sca12}, in particular by enhancing the direct two-proton decay \cite{mar12}. 
The collision term is known to be responsible for the additional spreading width of giant resonances as well as a fragmentation of 
their strength function~\cite{Lac04}.
They are also expected to modify the structure of the spectra of emitted nucleons by coupling 1p1h states to 2p2h. 
A first step would be to use the Extended-TDHF \cite{Won78,Ayi80,Lac99} or the time-dependent Density-Matrix~\cite{Wan85,Toh01,Ass09} 
formalisms. Because of computational limitations, spherical symmetry (and thus, only monopole vibrations) might be first considered.
Realistic spectra of emitted particles could then be computed and compared to experimental data. 
However, the reconstruction of the strength functions from these spectra will be more complicated as the simple technique introduced here, i.e., introducing a shift of the single particle hole energy, is valid only for 1p1h contributions. 
Finally, one could analyze the hole structure of the remaining nucleus as well, using, e.g., similar techniques as in Ref.~\cite{sim10,sim11} to determine the particle number distributions for different $l$ and $j$ quantum numbers. 

\thanks{
We  thank T. Lesinski for providing  the routine for computation of Bessel functions.
We are also grateful for comments and discussions with Ph. Chomaz and D. Lacroix at the 
early stage of this work. 
Part of the calculations have been performed at the NCI National Facility in Canberra, Australia, which is supported by the Australian Commonwealth Government. 
Partial support from ARC Discovery grants DP0879679 and 
DP110102879, as well as ARC Future Fellowship FT120100760 is acknowledged.
}

\end{document}